\documentstyle[preprint,aps]{revtex}

\begin{document}
\draft
\title{A potential multipartide entanglement measure}
\author{Alexander Wong\cite{email} and Nelson Christensen \cite{email2}}
\address{Physics and Astronomy, Carleton College, Northfield, Minnesota, USA, 55057\
\
}
\date{\today}
\maketitle

\begin{abstract}
In this Brief Report we discuss entanglement of multiparticle quantum
systems. We propose a potential measure of a type of entanglement of pure
states of $n$ qubits, the $n$-tangle. For a system of two qubits the
n-tangle is equal to the square of the concurrence, and for systems of three
qubits it is equal to the ''residual entanglement''. We show that the $n$%
-tangle, is also equal to the generalization of concurrence squared for even 
$n$, and use this fact to prove that the $n$-tangle is an entanglement
monotone. However, the $n$-tangle is undefined for odd $n>3$. Finally we
propose a measure related to the $n$-tangle for mixed state systems of $n$
qubits, and find an analytical formula for this measure for even $n$.
\end{abstract}


\bigskip 

\narrowtext

\section{Introduction}

The quantum phenomenon of entanglement is presently a subject of much active
research and discussion. This comes from fundamental interest in quantum
phenomena, and is also due to recent proposals for quantum computation \cite
{qcomp,qcomp2}. Entanglement is the property that provides a quantum
computer advantages over its classical counterpart. If one is designing a
quantum computer then quantifying the entanglement of a large number of
qubits is likely to be valuable. Quantum entanglement allows correlations
between separated quantum particles that are not possible in classical
systems \cite{EPR}. Hence, entanglement measures should also prove valuable
in the quantum applications of cloning, communication and encryption.

A method for classifying and quantifying the entanglement in a particular
state would greatly increase our understanding of this phenomenon; there
have been numerous studies into quantum entanglement, with equally numerable
entanglement measures proposed \cite{1,2,3,5,6,7,eisert,Acin,dur,chb,beg}.
There remain many open questions regarding quantification of entanglement.
In particular, states with more than one subsystem have only just begun to
be considered. While entanglement measures of pure states are essential, so
is their applicability to mixed states. The presence of noise in a quantum
channel \cite{noise}\ or the decoherence effects of qubits interacting with
an environment \cite{deco} will transform an idealized pure state into a
mixed one.

One type of multipartide entanglement is $n$-way or $n$-party entanglement,
entanglement that critically involves all $n$ particles. \ For example, a
three qubit state with only three-way (or three-party) entanglement has the
property that tracing out one of the qubits leaves the other two particles
unentangled \cite{5} It was recently proven that states with $n$-way
entanglement ($n>2$) cannot be reversibly distilled from two-way
entanglement \cite{7}. \ An example of a state with only three-way
entanglement is the $GHZ$ state: $|GHZ>=(|000>+\ |111>)/\sqrt{2}$, for which
case $\tau _{ABC}(|GHZ>)=1$. The $W$ state, $|W>=(|001>+\ |010>+\ |100>)/%
\sqrt{3}$, with $\tau _{ABC}(|W>)=0,$ is an example of a state with two-way
but no three-way entanglement: tracing out one of the particles leaves a
partially entangled pair of qubits. In general, three-qubit states have both
kinds of entanglement.

The {\em concurrence} has been shown to be a useful entanglement measure for
pure and mixed states with two qubits, and can be related to the {\em %
entanglement of formation} \cite{2}. A recent paper by Coffman, Kundu, and
Wootters \cite{1} using concurrence to examine three qubit quantum systems
introduced the concept of ``residual entanglement'', or the 3 - tangle, $%
\tau _{ABC}$. $\tau _{ABC}(|\psi >)$ is a potential way to quantify the
amount of three-way entanglement in the system ABC.

In this Brief Report we will show that a generalization of the $3$-tangle
for $n$ qubits, the $n$-tangle $\tau $, is related to a generalization of
pure state concurrence for states with an even number of qubits. This allows
us to prove that the $n$-tangle is an entanglement monotone for states with
three or an even number of qubits. We also show that the $n$-tangle equals $%
1 $ for the $n$ qubit generalization of {\em GHZ} state \cite{GHZ}, and $0$
for the $n$ qubit generalization of the {\em W} state \cite{5}. Lastly, we
introduce a mixed state measure of entanglement related to $n$-tangle that
is analogous to the entanglement of formation and find an analytical formula
for this measure for states with an even number of qubits.

The Brief Report is organized as follows. In Sec. II we define the $n$%
-tangle and show that for states with even $n$, $\tau _{1...n}$ is equal to
the square of a natural generalization of pure state concurrence. Since two
qubit concurrence is related to entanglement and entanglement of formation 
\cite{2}, this suggests that $n$-tangle may have a physical interpretation.
We prove that $\tau _{1...n}$ is an entanglement monotone \cite{3}, which
gives further evidence that the $n$-tangle measures a type of entanglement.
We also consider the value of $n$-tangle for generalizations of the {\em GHZ}
and {\em W} states and another example state. The extension of our pure
state results to mixed states is shown in Sec. III. A mixed state version of
the $n$-tangle, $\tau ^{min}$, is introduced, and an analytical formula for $%
\tau _{1...n}^{min}$ for even $n$ is presented. In Sec. IV we conclude with
a discussion of our results.

\section{The $n$-tangle}

For three qubits the ''residual entanglement'', or $\tau _{ABC}$, is given by

$$
\tau _{ABC}(|\psi >)=2\left| {\sum a_{\alpha _{1}\alpha _{2}\alpha
_{3}}a_{\beta _{1}\beta _{2}\beta _{3}}a_{\gamma _{1}\gamma _{2}\gamma
_{3}}a_{\delta _{1}\delta _{2}\delta _{3}}\epsilon _{\alpha _{1}\beta
_{1}}\epsilon _{\alpha _{2}\beta _{2}}\epsilon _{\gamma _{1}\delta
_{1}}\epsilon _{\gamma _{2}\delta _{2}}\epsilon _{\alpha _{3}\gamma
_{3}}\epsilon _{\beta _{3}\delta _{3}}}\right| \eqno(1) 
$$
where the $a$ terms are the coefficients in the standard basis defined by

\noindent $|\psi >=\sum_{i_{1}...i_{n}}a_{i_{1}...i_{n}}|i_{1}i_{2}...i_{n}>$
and $\epsilon _{01}=-\epsilon _{10}=1$ and $\epsilon _{00}=-\epsilon _{11}=0$
\cite{1}. We define the $n$-tangle by

$$
\tau _{1...n}=2\left| {\sum a_{\alpha _{1}...\alpha _{n}}a_{\beta
_{1}...\beta _{n}}a_{\gamma _{1}...\gamma _{n}}a_{\delta _{1}...\delta
_{n}}\epsilon _{\alpha _{1}\beta _{1}}\epsilon _{\alpha _{2}\beta
_{2}}...\epsilon _{\alpha _{n-1}\beta _{n-1}}\epsilon _{\gamma _{1}\delta
_{1}}\epsilon _{\gamma _{2}\delta _{2}}...\epsilon _{\gamma _{n-1}\delta
_{n-1}}\epsilon _{\alpha _{n}\gamma _{n}}\epsilon _{\beta _{n}\delta _{n}}}%
\right| \eqno(2) 
$$
for all even $n$ and $n=3$. By reasoning similar to that used for $n=3$ \cite
{5}, the $n$-tangle is invariant under local unitarities. We show below that
the $n$-tangle is invariant under permutations of the qubits. However, the
above formula is {\it not }invariant under permutations of qubits for
general odd $n$ over 3, and hence is not a viable measure of odd-way
entanglement (aside from $n=3$).

There is a relationship that can be shown between $\tau $ and pure state
concurrence. Pure state concurrence is defined for states of two qubits in 
\cite{2} by $C(\psi )=|<\psi |\tilde{\psi}>|^{2}$, where $|\tilde{\psi}%
>=\sigma _{y}^{\otimes n}|\psi ^{\ast }>$ is the ``spin flip'' of $|\psi >$
in terms of the Pauli spin matrix $\sigma _{y}=\left( 
\begin{array}{cc}
0 & -i \\ 
i & 0
\end{array}
\right) $. $C$ is defined only for states of two qubits, but the obvious
generalization uses the same equation, $C_{1...n}(\psi )=|<\psi |\tilde{\psi}%
>|^{2}$, where $|\tilde{\psi}>$ now is for an $n$-qubit state. Note that for
the two qubit case, $\tau _{12}=C^{2}$. We will prove that the analogous
equation, $\tau _{1...n}=C_{1...n}^{2}$ is true for all even $n$.

We shall find an expression for $C_{1...n}^{2}$ in terms of the coefficients
in the standard basis. One can express an $n$ qubit state $|\psi >$ as a
vector in the standard basis indexed by $|\psi >_{i_{1}...i_{n}}$, where
each $i$ indexes one of the qubits. Then $|\psi
>_{i_{1}...i_{n}}=a_{i_{1}...i_{n}}$.

\vskip12 pt Note that $\sigma _{y_{i_{1}...i_{n},j_{1}...j_{n}}}^{\otimes
n}=\epsilon _{i_{1}j_{1}}...\epsilon _{i_{n}j_{n}}\ast e^{i\theta }$ for
some real $\theta $ because $\sigma _{y_{i,j}}=-i\ast \epsilon _{ij}$.
Therefore, $|\tilde{\psi}>=\sigma _{y}^{\otimes n}|\psi ^{\ast }>$ implies $|%
\tilde{\psi}>_{i_{1}...i_{n}}=\sum_{\beta _{1}...\beta _{n}}^{1}a_{\beta
_{1}...\beta _{n}}^{\ast }\epsilon _{i_{1}\beta _{1}}\epsilon _{i_{2}\beta
_{2}}...\epsilon _{i_{n}\beta _{n}}\ast e^{i\theta }$ so

\noindent $<\psi |\tilde{\psi}>=\sum_{all\ \alpha ,\beta }a_{\alpha
_{1}...\alpha _{n}}^{\ast }a_{\beta _{1}...\beta _{n}}^{\ast }\epsilon
_{\alpha _{1}\beta _{1}}\epsilon _{\alpha _{2}\beta _{2}}...\epsilon
_{\alpha _{n}\beta _{n}}\ast e^{i\theta }$. Thus,

$$
|<\psi |\tilde{\psi}>|^{2}=\left| {\sum a_{\alpha _{1}...\alpha
_{n}}a_{\beta _{1}...\beta _{n}}a_{\gamma _{1}...\gamma _{n}}a_{\delta
_{1}...\delta _{n}}\epsilon _{\alpha _{1}\beta _{1}}\epsilon _{\alpha
_{2}\beta _{2}}...\epsilon _{\alpha _{n}\beta _{n}}\epsilon _{\gamma
_{1}\delta _{1}}\epsilon _{\gamma _{2}\delta _{2}}...\epsilon _{\gamma
_{n}\delta _{n}}}\right| \eqno
(3) 
$$
where the sum is over all indices. Expanding the last index of each $a$ and
using the fact that $\epsilon _{i,j}=-\epsilon _{j,i}$ for even $n$, one gets

\begin{eqnarray}
| &<&\psi |\tilde{\psi}>|^{2}=  \nonumber \\
&&2|-\sum a_{\alpha _{1}...\alpha _{n-1}0}\ a_{\beta _{1}...\beta _{n-1}1}\
a_{\gamma _{1}...\gamma _{n-1}1}\ a_{\delta _{1}...\delta _{n-1}0}\epsilon
_{\alpha _{1}\beta _{1}}\epsilon _{\alpha _{2}\beta _{2}}...\epsilon
_{\alpha _{n-1}\beta _{n-1}}\epsilon _{\gamma _{1}\delta _{1}}\epsilon
_{\gamma _{2}\delta _{2}}...\epsilon _{\gamma _{n-1}\delta _{n-1}}  \nonumber
\\
&&-\sum a_{\alpha _{1}...\alpha _{n-1}1}\ a_{\beta _{1}...\beta _{n-1}0}\
a_{\gamma _{1}...\gamma _{n-1}0}\ a_{\delta _{1}...\delta _{n-1}1}\epsilon
_{\alpha _{1}\beta _{1}}\epsilon _{\alpha _{2}\beta _{2}}...\epsilon
_{\alpha _{n-1}\beta _{n-1}}\epsilon _{\gamma _{1}\delta _{1}}\epsilon
_{\gamma _{2}\delta _{2}}...\epsilon _{\gamma _{n-1}\delta _{n-1}}| 
\eqnum{4}
\end{eqnarray}

\noindent Now we turn our attention to the expression for $\tau $. Eq. (2)
can be expanded to

\begin{eqnarray}
\tau _{1...n} &=&|\sum a_{\alpha _{1}...\alpha _{n-1}0}\ a_{\beta
_{1}...\beta _{n-1}0}\ a_{\gamma _{1}...\gamma _{n-1}1}\ a_{\delta
_{1}...\delta _{n-1}1}\epsilon _{\alpha _{1}\beta _{1}}\epsilon _{\alpha
_{2}\beta _{2}}...\epsilon _{\alpha _{n-1}\beta _{n-1}}\epsilon _{\gamma
_{1}\delta _{1}}\epsilon _{\gamma _{2}\delta _{2}}...\epsilon _{\gamma
_{n-1}\delta _{n-1}}  \nonumber \\
&&+\sum a_{\alpha _{1}...\alpha _{n-1}1}\ a_{\beta _{1}...\beta _{n-1}1}\
a_{\gamma _{1}...\gamma _{n-1}0}\ a_{\delta _{1}...\delta _{n-1}0}\epsilon
_{\alpha _{1}\beta _{1}}\epsilon _{\alpha _{2}\beta _{2}}...\epsilon
_{\alpha _{n-1}\beta _{n-1}}\epsilon _{\gamma _{1}\delta _{1}}\epsilon
_{\gamma _{2}\delta _{2}}...\epsilon _{\gamma _{n-1}\delta _{n-1}}  \nonumber
\\
&&-\sum a_{\alpha _{1}...\alpha _{n-1}0}\ a_{\beta _{1}...\beta _{n-1}1}\
a_{\gamma _{1}...\gamma _{n-1}1}\ a_{\delta _{1}...\delta _{n-1}0}\epsilon
_{\alpha _{1}\beta _{1}}\epsilon _{\alpha _{2}\beta _{2}}...\epsilon
_{\alpha _{n-1}\beta _{n-1}}\epsilon _{\gamma _{1}\delta _{1}}\epsilon
_{\gamma _{2}\delta _{2}}...\epsilon _{\gamma _{n-1}\delta _{n-1}}  \nonumber
\\
&&-\sum a_{\alpha _{1}...\alpha _{n-1}1}\ a_{\beta _{1}...\beta _{n-1}0}\
a_{\gamma _{1}...\gamma _{n-1}0}\ a_{\delta _{1}...\delta _{n-1}1}\epsilon
_{\alpha _{1}\beta _{1}}\epsilon _{\alpha _{2}\beta _{2}}...\epsilon
_{\alpha _{n-1}\beta _{n-1}}\epsilon _{\gamma _{1}\delta _{1}}\epsilon
_{\gamma _{2}\delta _{2}}...\epsilon _{\gamma _{n-1}\delta _{n-1}}| 
\nonumber \\
&&  \eqnum{5}
\end{eqnarray}
Consider some term in the fully expanded version of the first line of the
above equation, $a_{\mu _{1}...\mu _{n-1}0}\ a_{\bar{\mu}_{1}...\bar{\mu}%
_{n-1}0}\ a_{\nu _{1}...\nu _{n-1}1}\ a_{\bar{\nu}_{1}...\bar{\nu}_{n-1}1}$,
where $\bar{\mu}=1\ \hbox{if}\ \mu =0\ \hbox{and}\ \bar{\mu}=0\ \hbox{if}\
\mu =1$. This term can be positive or negative. The expansion of the first
line of the above equation also contains the term

\noindent $a_{\bar{\mu}_{1}...\bar{\mu}_{n-1}0}\ a_{\mu _{1}...\mu _{n-1}0}\
a_{\nu _{1}...\nu _{n-1}1}\ a_{\bar{\nu}_{1}...\bar{\nu}_{n-1}1}$. For even $%
n$ the sign of this term will be opposite the sign of the original term
since the signs of an odd number of $\epsilon $'s have been flipped. So the
two above terms will add to zero, as will all other terms in the first line
of Eq. (5). The second line of Eq. (5) also goes to zero by the same
argument. Thus, $\tau _{1...n}=|<\psi |\tilde{\psi}>|^{2}$ for all even $n$.

This equality indicates that $n$-tangle is a more natural measure of
entanglement than concurrence because, for odd $n$, $C_{1...n}=0$, while a
meaning of $n$-tangle is already established for $n$ = 3 \cite{1}. From Eq.
(3) one can determine that the quantity $C_{1...n}^{2}=\tau _{1...n}$ for
even $n$ is invariant under permutations of the qubits, since changing the
order of the indices (i.e. the numbering of the Greek letters) is only
renaming the indices. This allows us to apply the method used in \cite{5} to
prove that $\tau _{ABC}$ is an entanglement monotone to prove that the $n$%
-tangle is an entanglement monotone, a property that good measures of
entanglement must satisfy \cite{3}. As in \cite{5} (we explicitly follow
their form and proof outline), the invariance of the $n$-tangle under
permutations of the parties lets us consider local positive operator valued
measures (POVM's) for one party only. Let $A_{1},A_{2}$ be two POVM elements
such that $A_{1}^{\dagger }A_{1}+A_{2}^{\dagger }A_{2}=I$, then $%
A_{i}=U_{i}D_{i}V$ with $U_{i}$ and $V$ being unitary matrices, and $D_{i}$
being diagonal matrices with entries $(a,b)$ and $(\sqrt{1-a^{2}},\sqrt{%
1-b^{2}})$ respectively. For some initial state $|\psi >$ let $|\hat{\phi}%
_{i}>=A_{i}|\psi >$ be the subnormalized states obtained after application
of the POVM. Let $|\phi _{i}>=|\hat{\phi}_{i}>/\sqrt{p_{i}}$, $p_{i}=<\hat{%
\phi}_{i}|\hat{\phi}_{i}>$. Then 
$$
<\tau >=p_{1}\tau (\phi _{1})+p_{2}\tau (\phi _{2})\eqno
(6) 
$$
Since the $n$-tangle is invariant under local unitarities \cite{5} $\tau
(U_{i}D_{i}V\psi )=\tau (D_{i}V\psi )$. Now, noting that every term of Eq.
(2) contains two $a$'s with subscripts starting with zeros and two $a$'s
with subscripts starting with ones and that every term is quartic with
respect to the $a$'s, it can be shown that 
$$
\tau (\phi _{1})={\frac{a^{2}b^{2}}{p_{1}^{2}}}\tau (\psi ),\tau (\phi _{2})=%
{\frac{(1-a^{2})^{2}(1-b^{2})^{2}}{p_{2}^{2}}}\tau (\psi )\eqno
(7) 
$$
Defining $P_{0}$ to be the sum of the squared magnitudes of the first $%
2^{n-1}$ components of $|\psi >$ in the standard basis, and $P_{1}$ to be
the sum of the squared magnitudes of the last $2^{n-1}$ components of $|\psi
>$, we can say 
$$
p_{1}=a^{2}P_{0}+b^{2}P_{1}\ \hbox{and}\ p_{2}=(1-a^{2})P_{0}+(1-b^{2})P_{1}%
\eqno
(8) 
$$
Combining Eqs. (6-8) with the fact that $P_{0}+P_{1}=1$ some algebra shows
that $<\tau >/\tau (\psi )\leq 1$ thus proving that $n$-tangle is an
entanglement monotone.

Some examples provide further support for the $n$-tangle being a measure of
some type of $n$-party entanglement. An $n$-qubit CAT state, $(|0^{\otimes
n}>+|1^{\otimes n}>)/\sqrt{2}$ is a state with entirely $n$-way
entanglement; measuring any one of the qubits in the standard basis
determines the value of all of the other qubits, however, if one of the
qubits is traced out, the remaining qubits are unentangled. For these
states, the $n$-tangle is 1; all terms in Eq. (2) go are 0 for an $n$-CAT
state except for when $\alpha _{i}=0,\beta _{i}=1,\gamma _{i}=1\ \hbox{and}$ 
$\delta _{i}=0\ \hbox{or}$ $\alpha _{i}=1,\beta _{i}=0,\gamma _{i}=0\ %
\hbox{and}$ $\delta _{i}=1$, so $\tau _{1...n}(|CAT>)=2|-1/4+-1/4|=1$.
Another interesting set of states are the $n$-qubit $W$ states \cite{5} $%
(|0...01>+|0...010>+...+|10...>)/\sqrt{n}.$ \ For these states, the equality 
\cite{1} 
$$
C_{12}^{2}+C_{13}^{2}+...+C_{1n}^{2}=C_{1(23...n)}^{2}\eqno
(9) 
$$
holds, thus tracing out all but two of the qubits leaves the two remaining
qubits partially entangled. Note that $C_{1(23...n)}^{2}\not=0$ and that the
W states are symmetric. Note that the $\epsilon $'s assure that all terms in
Eq. (2) are $0$ for $W$ states, and hence $n$-tangle is zero for W states
(except for $n=2$, where $\tau =1$, since $\tau $ measures two way
entanglement in this case). From the above examples, it is tempting to
hypothesize that $n$-tangle is a measure of $n$-way entanglement, but a
counterexample shows otherwise: Consider the four qubit pure state that is
the tensor product of two singlet states. A simple calculation shows that $4$%
-tangle has a value of 1 for this state. If $4$-tangle measured $4$-way
entanglement, its value should have been 0, since this state has no
entanglement between the pairs of entangled qubits. Thus, while $n$-tangle
appears to be related to some kind of multipartide entanglement, it is not
by itself a measure of $n$-way entanglement.

\section{Mixed state generalization of $n$-tangle}

We would like to have a mixed state generalization of $n$-tangle. Such a
quantity would enable us to classify and quantify even more types of
entanglement. For example, a four qubit pure state would have $6$ values of
the mixed state $2$-tangle between each of the pairs of qubits, $4$ values
for the mixed state $3$-tangle between each set of $3$ qubits, and a value
for the $4$-tangle. We suggest defining, for an $n$-qubit mixed state $\rho $%
, $\tau ^{min}(\rho )$ to be the minimum of $\sum_{i}p_{i}\tau (\psi _{i})$
for all pure state decompositions of $\rho $, given by $\rho
=\sum_{i}p_{i}|\psi ><\psi |$. This is analogous to the entanglement of
formation \cite{6}, and in fact the entanglement of formation is a function
of $\tau ^{min}(\rho _{12})$ for states of two qubits \cite{2}. This
definition is also justified by the fact that Eq. (24) of \cite{1} can now
be rewritten as 
$$
C_{A(BC)}^{2}=\tau _{AB}^{min}+\tau _{AC}^{min}+\tau _{ABC}\eqno
(10) 
$$
That is, for two qubits, $\tau _{12}^{min}(\rho )$ already has physical
significance, so it appears to be a natural way to define a mixed state $%
\tau $. Now, in \cite{2}, Wootters presents a proof that $C_{min}(\rho
)=max\{0,\lambda _{1}-\lambda _{2}-\lambda _{3}-\lambda _{4}\}$ where $%
\lambda _{i}$ is the square root of the $i^{th}$ eigenvalue, in decreasing
order, of $\rho \tilde{\rho}$. This proof is generalizable to show that $%
C_{min}(\rho )=max\{0,\lambda _{1}-\lambda _{2}...-\lambda _{n}\}$ for an $n$%
-qubit system, and therefore $\tau ^{min}\left( \rho \right)
=C_{min}^{2}\left( \rho \right) =\left[ max\{0,\lambda _{1}-\lambda
_{2}...-\lambda _{n}\}\right] ^{2}$. \ This result is also a subset of a
more general proof by Uhlmann \cite{4}.

A large number of doubts remain about the meaning of $n$-tangle. In
particular, we would like to have a physically meaningful definition of $n$%
-way entanglement so that we could compare $n$-tangle and other multipartide
entanglement measures with meaningful values. It seems likely that $n$%
-tangle, in combination with other multipartide entanglement measures (most
likely the $n$-tangles of smaller subsystems within a given state) will be
related to a multipartide generalization of the $2$-qubit entanglement, $E$,
related to the Shannon entropy. Unfortunately, no such generalization of
entanglement is obvious. If a formula for $\tau _{1...n}^{min}$ for $n=3$
could be found, it might be possible to prove statements analogous to Eq.
(10) which would lend more legitimacy to $n$-tangle. We would also like to
have a generalization of $n$-tangle for states with subsystems larger than
qubits.

\section{Discussion}

In summary, we have proposed a potential measure of a type of $n$-partide
entanglement of pure and mixed states: for pure states, the $n$-tangle, and
for mixed states the related $\tau _{1..n}^{min}$. These measures show many
signs of being useful ways to quantify a type of multipartide entanglement.
For even $n$, $n$-tangle and $\tau _{1..n}^{min}$ are equal to the square of
a generalization of pure and mixed state concurrence, $n$-tangle is also an
entanglement monotone. $n$-tangle has values of $1$ for $n-CAT$ states and
values of $0$ for $W$ states where $n>2$ but has a value of 1 for a product
state of two singlets. Hopefully these measures will further our
understanding of multi-partide entanglement. In particular, further
exploration of their mixed state forms may lead to the discovery of
relationships between different types of entanglement within a particular
system.

\acknowledgments
We would like to thank W.K. Wootters for a great deal of useful advice. This
work was supported by the Kresge Foundation and Carleton College.

\appendix

\end{document}